# MAXILLOFACIAL COMPUTER AIDED SURGERY: A 5 YEARS EXPERIENCE AND FUTURE

CH. MARECAUX[1,2], M. CHABANAS[1], V. LUBOZ[1,3], A. PEDRONO[3], F. CHOULY[1], P. SWIDER[3], Y. PAYAN[1], F. BOUTAULT[2].

1. TIMC-GMCAO laboratory, UMR CNRS 5525, Faculté de Médecine Domaine de la Merci, 38706 La Tronche, France
{Chistophe.Marecaux, Matthieu.Chabanas, Vincent.Luboz, Yohan.Payan}@imag.fr
2. INSERM, Biomecanic, CHU Purpan, BP 3103, 31026 Toulouse Cedex 3, France
{apedrono, pascal.swider}@toulouse.inserm.fr
3. Department of Maxillofacial and Facial Plastic Surgery, CHU Purpan, 31059 Toulouse Cedex, France
Boutault.F@chu-toulouse.fr

Since early eighties, cranio-maxillofacial surgery was interested in computer aided techniques. It was first as an aid for diagnostic. CT imaging allowed skull 3D reconstruction and anatomic visualization of craniofacial malformations close from surgical experience [1].

Then, in the nineties, surgical navigation techniques were adapted from neurosurgery to maxillofacial surgery[2].

The department of maxillofacial and facial plastic surgery from Purpan Hospital in Toulouse has had a computer aided navigation system for five years, in collaboration with departments of neurosurgery and ENT surgery. Its current utilization is quietly limited to strict navigation by lack of the specific tools for cranio-maxillofacial surgery. By the way, we try to extend these applications: we present some of them and specially in a specific pathology: orbital surgery for exophthalmia.

According to us, an increased utilization of computer aided techniques in cranio-maxillofacial surgery requires specific softwares for planning and simulation on one part and intra operative navigation and guidance on other part. We address planning and simulation in two pathologies: orthognatic surgery and orbital surgery in exophthalmia for Basedow illness.

## A 5 YEARS EXPERIENCE IN COMPUTER AIDED NAVIGATION

Computer aided navigation with the Stealth Station[©] (Sofamor Danek) and the Xomed software[©] is possible as previously described in literature [3,4]. Tools (a reference framework and a pointer) are located in space by 3 active diodes each, infrared emitting. These active tools are seen by an optical tracker (a camera). At the beginning of the surgical operation, the patient's anatomy and the virtual patient's image reconstructed from the pre operative CT scan imaging are rigidly matched point for point. Then, the pointer tip is located in the surgical field and showed in the same time on the CT scan or NMR slides. Planning is limited to surgical trajectories between two defined points.

### *1. Utilization of a navigation system in cranio-maxillofacial surgery: examples*

We use a system for intra operative localization in surgeries located in difficult areas because of important anatomic element proximity or a poor overview by a small surgical access:
- Surgery for cancer located in orbit or skull base for example.
- Orbital surgery as an aid for setting up a bone graft or an implant, or for optimising decompression osteotomy.

Using the "surgical planning" function, we have "extended" applications for:
- Wiring implant in a small bone volume (implant, mandibular distractor).
- Zygomatic osteotomy repositioning.

### *2. About orbital surgery for exophtalmia*

Exophthalmia in Basedow disease is an excessive forward displacement of the ocular globe outside the orbit, by orbital fat or muscle volume increasing.

Its consequences can be functional (cornea injury, diplopia, lost of visual acuity) or aesthetic.

Surgery can be used for exophthalmia reduction by removing inferior, medial or lateral orbital walls, according to the required backward displacement.

Computer aided navigation is used for optimising orbital osteotomy, preserving the optical nerve.

## PLANNING AND SIMULATION: BASIS AND EXAMPLES IN CRANIO-MAXILLOFACIAL SURGERY

Computer aided maxillofacial surgery has been well defined in literature [5,6,7] and can be summarized as on figure 1.

We address planning and simulation stages in orthognatic surgery and orbital surgery for exophthalmia.

Surgical planning definition requires first to integrate patient's data for diagnostic, then to simulate different therapeutic options to choose the more relevant. It needs statistical, anatomical, physiological or biomechanical models for studied organs or pathologies.

### *1   Orthognatic surgery*

### *1.1   Purpose*

Orthognatic surgery attempts to establish normal aesthetic and functional anatomy for patients suffering from dentofacial disharmony. In this way, current surgery aims at normalizing patients dental occlusion, temporo-mandibular joint function and morphologic appearance by repositioning maxillary and mandibular skeletal osteotomized segments. Soft tissue changes are mainly an adaptation of these bones modifications.

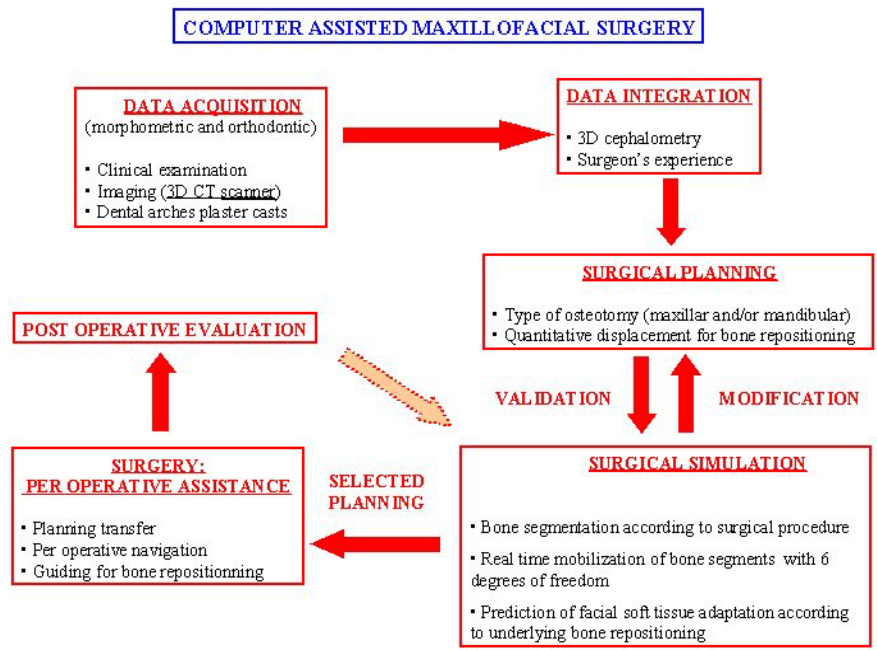

Figure 1: Principles for computer aided cranio-maxillofacial surgery

In current practice, the orthognatic surgical planning involves several multimodal data: standard radiographies for bidimensional cephalometric analysis, plaster dental casts, photographs and clinical evaluation. This difficult and laborious process might be responsible of imprecision and requires a strong experience. Moreover, there is no relevant way to predict aesthetic postoperative outcomes.

Computer assisted surgical technologies may improve current orthognatic protocol as an aid in diagnostic and surgical planning. It requires a 3D method for cephalometric analysis and a model for facial soft tissue adaptation.

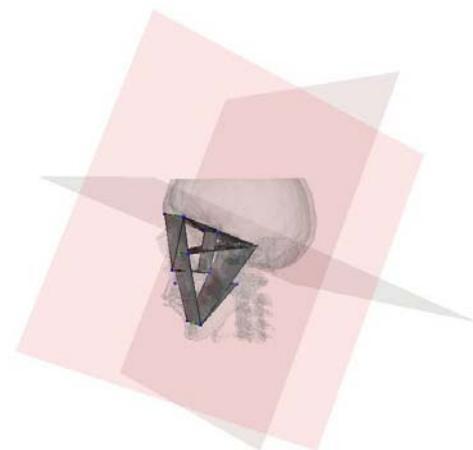

Figure 2: Example of 3D cephalometric analysis.

## 1.2 Method

The skeletal analysis (3D cephalometry) is composed by a maxillofacial frame [8], built from anatomic landmarks (figure 2). Mathematical tools allow metric, angular and surfacic measurements.

The biomechanical facial soft-tissue model uses finite element (FE) method [9]. First, a "generic" model of the face is manually built, integrating skin layers and muscles. Then, the mesh of this generic model is conformed to each patient morphology, using an elastic registration method and patient data segmented from CT images (figure 3). This automatically generated specific model is suitable for simulation by displacing internal nodes as the surgical procedure does.

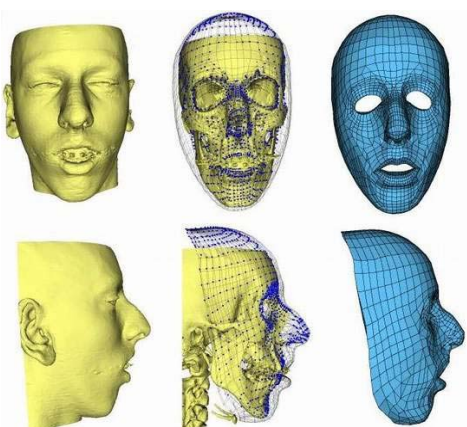

Figure 3: Example of specific patient's facial soft-tissue model.

### 1.3 Discussion

These models support the surgical bone osteotomies simulation and the post operative facial soft-tissues deformation prediction. First clinical validations have been done (figure 4).

There are actually few propositions for 3D cephalometric analysis [10,11]. They are mainly orthodontic analysis oriented and unusable for computer aided surgery.

These models are friendly and a few time consuming, usable for current practice. Nevertheless, more developments and precisions are still required.

Intra operative guidance will be required for respect of the relevant planning.

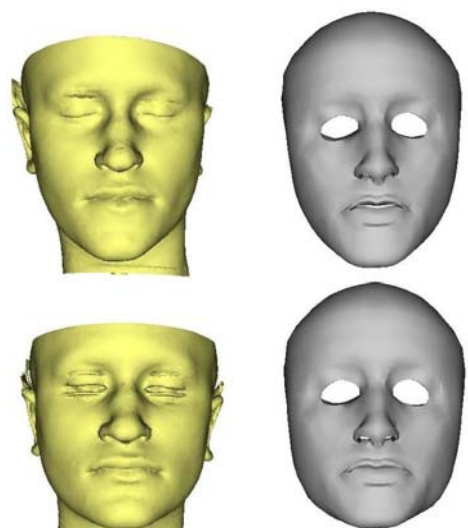

Figure 4: Example of post operative prediction compared to surgical outcomes

## 2 Surgery for exophthalmia

### 2.1 Purpose

Current therapeutic planning in surgery for exophthalmia is quite empiric about localization and size of the orbital wall osteotomy. Planning respect is mostly important in this bilateral surgery.

Planning might be improved by an orbit model for backward displacement prediction according to the osteotomy. Intra operative navigation will be helpful for strict respect of the relevant planning.

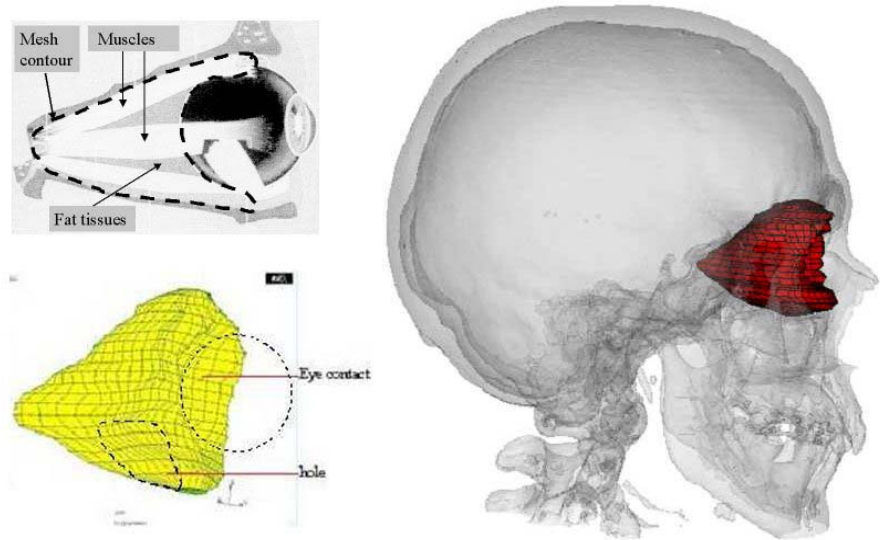

Figure 5: Orbital 3D FE generic mesh.

## 2.2 Method

A biomechanical 3D Finite Element (FE) model of the orbit [12], including muscles, fat tissues and orbital walls is defined. First, a generic 3D model is built. Then, the generic mesh is adapted to specific patient's anatomy segmented from CT scan imaging by the mean of an elastic registration. Surgical procedure is simulated by changing boundary conditions according to osteotomy size and localization, and imposing a pressure on the ocular globe surface.

## 2.3 Discussion

There is no such previous FE orbital model. Its qualitative relevance for orbital decompression simulation has been shown. Neitherless, quantitative validation has to be done.
This method is still time computing consuming and has to be improved for current clinical application.

## Conclusion

Computer aided techniques improve surgical performance. Its use will be mostly increased by developing specific models and softwares for the different studied pathologies. After planning and simulation, we will have to address navigation.